\documentclass{jfm}

\usepackage{graphicx}
\usepackage{natbib}

\ifCUPmtlplainloaded \else
  \checkfont{eurm10}
  \iffontfound
    \IfFileExists{upmath.sty}
      {\typeout{^^JFound AMS Euler Roman fonts on the system,
                   using the 'upmath' package.^^J}%
       \usepackage{upmath}}
      {\typeout{^^JFound AMS Euler Roman fonts on the system, but you
                   dont seem to have the}%
       \typeout{'upmath' package installed. JFM.cls can take advantage
                 of these fonts,^^Jif you use 'upmath' package.^^J}%
      }
  \else
  \fi
\fi

\ifCUPmtlplainloaded \else
  \checkfont{msam10}
  \iffontfound
    \IfFileExists{amssymb.sty}
      {\typeout{^^JFound AMS Symbol fonts on the system, using the
                'amssymb' package.^^J}%
       \usepackage{amssymb}%

      }{}
  \fi
\fi

\ifCUPmtlplainloaded \else
  \IfFileExists{amsbsy.sty}
    {\typeout{^^JFound the 'amsbsy' package on the system, using it.^^J}%
     \usepackage{amsbsy}}
    {\providecommand\boldsymbol[1]{\mbox{\boldmath $##1$}}}
\fi

\providecommand\bnabla{\boldsymbol{\nabla}}

\newcommand\Pen{\mbox{\textit{Pe}}}  

\newsavebox{\astrutbox}
\sbox{\astrutbox}{\rule[-5pt]{0pt}{20pt}}

\title[ Electromigration Dispersion]{Electromigration dispersion in  a capillary in the presence 
of electro-osmotic flow}
\author[S. Ghosal and Z. Chen]%
{S.\ls \ns G\ls H\ls O\ls S\ls A\ls L%
  \thanks{Email address for correspondence: s-ghosal@northwestern.edu},\ns
\and Z.\ns  C\ls H\ls E\ls N}

\affiliation{Department of Mechanical Engineering, Northwestern University, Evanston, IL 60208, USA}

\pubyear{2010}
\volume{650}
\pagerange{119--126}
\date{?; revised ?; accepted ?. - To be entered by editorial office}
\begin{document}

\maketitle

\begin{abstract}
The differential migration of ions in an applied electric field is the basis for separation of chemical species by 
capillary electrophoresis. Axial diffusion of the concentration peak limits
the separation efficiency. Electromigration dispersion is observed when the concentration
of sample ions is  comparable to that of the background ions. Under such conditions, 
the local electrical conductivity is significantly altered in the sample zone making the electric field, and therefore, the ion migration velocity 
concentration dependent. The resulting nonlinear wave exhibits shock like features, 
and, under certain simplifying assumptions, is described by Burgers' equation (S. Ghosal and Z. Chen {\it Bull. Math. Biol.} 2010 {\bf 72},  pg. 2047).
In this paper, we consider the more general  situation where the walls of the separation channel may have a non-zero zeta potential and are therefore 
able to sustain an electro-osmotic bulk flow. The main result is a one dimensional nonlinear advection diffusion equation for the area averaged 
concentration. This homogenized equation accounts for the Taylor-Aris dispersion resulting from  the  variation in the electro-osmotic slip velocity 
along the wall. It is shown that in a certain range of parameters,  the electro-osmotic flow can actually reduce the total dispersion 
by delaying the formation of a concentration shock. However, if the electro-osmotic flow is sufficiently high, the total dispersion 
is increased because of the Taylor-Aris contribution.
\end{abstract}


\section{Introduction}

Macromolecules such as DNA and proteins carry an electrical  charge in aqueous solution because of 
ionic dissociation of molecular groups on their surface. The amount and sign of the charge depends on various 
factors such as the pH of the solution. If an external electric field is applied, these macro-ions migrate along the 
field with a velocity that depends on its charge, size and shape as well as on the ionic composition of 
the background electrolyte. Electrophoresis is a technique of  analytical chemistry for separating a mixture of 
chemical compounds by exploiting their differing migration velocities in an applied electric field. It is a widely used 
laboratory technique in areas such as  molecular biology, forensics and medicine. 

The most common format is ``gel electrophoresis,''  where the electrolyte is suspended in a porous gel. The technique of capillary 
electrophoresis (CE) has developed rapidly in recent years, partly because of the possibility of integrating CE channels 
on to micro-fluidic chips. In CE, the sample and suspending electrolyte 
are contained in a micro-fluidic channel with width in the range of tens of microns. CE can proceed in several modes, the 
simplest of which is ``zone electrophoresis''.  Here a sample zone or band is injected into a micro-channel, which then moves 
by electrophoresis towards the detector at the opposite end under an applied electric field. Different components arrive at different times and their arrival
is marked by a peak in some solute property (usually the UV absorbance) at the detector window  \citep{czebook1,czebook2}.
The walls of the capillary, which are most commonly made of fused silica,  has an electrostatic charge that is characterized by its zeta potential.
Therefore, the applied electric field results in an electro-osmotic flow along the capillary \citep{probstein}. This flow is normally 
advantageous in CE because it sweeps both positive and negative ions past a single detector near the outlet and  also reduces the 
transit time from inlet to detector.
However, it can, under certain circumstances adversely affect the separation by causing ``anomalous dispersion''
of the sample peak \citep{ghosal_annrev06}. 

The physical process of separation in CE is the result of the mutually opposed and competing processes of differential 
migration of ions and diffusive spreading along the axis of separation. The resolution depends on the strength of 
the applied electric field. In practice, the applied voltage is often very large, in the range of kilo volts. The highest voltage 
that can be applied is limited by the ability to dissipate the considerable Joule heat that is produced in the buffer by the 
electrolytic current. This is the reason why the channel width may not exceed some tens of microns. It is clearly advantageous to have an
electrolytic buffer of low conductivity in order to minimize heating. It is also advantageous to have a relatively high concentration 
of sample ions, since one of the limitations of CE is the high demand placed on the sensitivity of the detector,
which must be sensitive to light attenuation over an optical path length of only a few tens of microns. 
However, the ratio of sample to background ion concentration cannot be increased indefinitely; distortions due to
 ``electromigration dispersion'' or the ``sample overloading effect''  limits the highest ratio of 
sample to background ion concentration that may be used in CE.

The underlying physical mechanism of electromigration dispersion was elucidated by \citet{mi_ev_ve_79a}  and may be 
roughly explained as follows: when the concentration of sample ions is comparable 
to that of the carrier electrolyte, the local electrical conductivity is altered in the vicinity of the sample peak. On the other hand, since the current
through the capillary must be the same at all points along the axis, 
the electric field must change locally. This follows since the product of the conductivity and electric field is the current (according to Ohm's law, neglecting 
for the time being the diffusion current due to ionic concentration gradients). This 
varying electric field alters the effective migration speed of the sample ions, which, in turn, alters 
its concentration distribution. Thus, we have  a nonlinear transport problem that must be solved in a self 
consistent manner. Electromigration dispersion causes highly asymmetric concentration profiles, high rates 
of dispersion and shock like structures that are reminiscent of nonlinear waves seen in many other physical systems. 
These effects have been widely reported in the literature on electrophoresis \citep{bouskova_elph04}.

Simple one dimensional mathematical models of electromigration dispersion have been studied by \citet{mi_ev_ve_79a,mikkers_ac99,math_th_elph_bk}
by invoking the assumption of vanishing  diffusivity which results in a single nonlinear hyperbolic equation for the concentration of sample ions. 
Solutions describe the  observed steepening of an intitially smooth profile leading to subsequent shock formation. Two recent reviews by \cite{gas09} and \cite{thormann09} provide more extensive 
references to this early work. The restriction of zero diffusivity was removed by \cite{ghosal_chen10} 
(henceforth referred to as GC) 
who considered a three ion system -- the sample ion, co-ion and counter-ion -- where the diffusivities of the three ionic species were equal 
but not necessarily zero.
The sample concentration was shown to obey a one dimensional nonlinear advection diffusion equation which reduced to Burgers' equation if the sample concentration was not too high 
relative to that of the background ions. Since the initial value problem for Burgers' equation may be exactly solved, useful insight into the nature of electromigration dispersion 
could be gained in the limits of small as well as large P\'{e}clet numbers. 
A generalization of this model to the situation where the buffer is a weak electrolyte has recently been presented~\citep{EMD1}.
It was found that the evolution of the peak shape may still be described by the Burgers' equation with a slight re-interpretation 
 of a certain model parameter.  If this parameter is taken as a fitting parameter, excellent quantitative 
agreement is obtained with measured values of the peak variance in experiments. 
A similar nonlinear advection diffusion equation was derived by \citet{zangle_propagation_2009}
in order to describe ``desalination shocks'' near constrictions in micro channels. However, in their problem the physical origin of the nonlinearity is related to 
surface conduction effects.

In this paper, we extend the earlier work of  GC by taking into account the fact that the channel wall generally has a non-zero zeta potential which 
creates a bulk fluid flow in the capillary due to electro-osmosis. The electro-osmotic flow affects the transport process in the following way: the
axial variation of the electric field created by the conductivity  changes results in a variable electro-osmotic slip velocity on the channel walls. This in turn results 
in an induced pressure gradient and the concomitant appearance of a shear in the velocity profile across the channel due to a well known mechanism \citep{he_00,gh_02b,gh_02a,gh_02c}. The shear  then enhances axial dispersion due to the 
Taylor-Aris effect \citep{taylor,aris}.  Our principal result is equation (\ref{generalized_GC}), which is a one dimensional 
nonlinear advection diffusion equation for the sample concentration averaged over the channel cross-section. 

The rest of the paper is organized as 
follows. In \S~\ref{sec:formulation} the complete mathematical statement of the flow and electro-diffusion problem is presented. In \S~\ref{sec:homogenized} a systematic reduction 
of the full equations leading to equation~(\ref{generalized_GC}) is achieved by introducing a number of physical approximations.  The consequences of this equation are discussed in \S~\ref{sec:dispersion},
where it is shown that there is a peak in the efficiency of separation at an intermediate value of the electro-osmotic flow strength. The broad 
implications of the analysis to separation efficiency in CE are discussed in \S~\ref{sec:conclusion}.

\section{Mathematical Formulation}
\label{sec:formulation}
\begin{figure}
  \centerline{\includegraphics[width=6in]{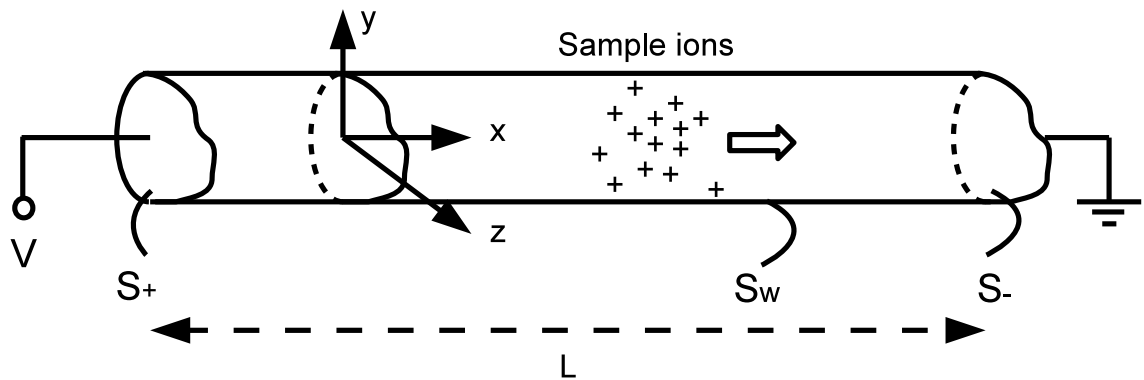}}
  \caption{Schematic diagram (not to scale) illustrating  the mathematical problem of electromigration 
  dispersion of a sample in CE. In addition to the sample ions shown, the capillary also contains a 
  carrier electrolyte consisting of co- and counter- ions.}
\label{fig:geom}
\end{figure}
We consider (see Figure~1) a uniform, long 
capillary of arbitrary cross-sectional shape connected to infinite reservoirs at either end. The length of the capillary is much greater than the 
width; the capillary may, for most purposes, be assumed infinitely long.
  The capillary contains a solute and $N$ strong (i.e. completely dissociated) electrolytes 
of concentration $c_i (x,y,z,t)$ where $i=1$ to $N$. These concentrations then obey the conservation equations based on the Nernst-Planck model for ion flux~\citep{probstein}
\begin{equation}
\frac{ \partial c_i}{\partial t} + \bnabla \cdot ( {\bf u} c_i - z_i e \nu_i c_i \bnabla \phi_e - D_i \bnabla c_i ) = 0,
\label{transport}
\end{equation} 
where $z_i$, $\nu_i$ and $D_i$ are the valence, mobility and diffusivity of species $i$,  the electronic charge is $e$, $\phi_e$ is the electric potential and 
${\bf u} (x,y,z,t)$ is the fluid flow velocity.
The potential $\phi_e$ obeys the Poisson equation (in CGS units)
\begin{equation} 
\varepsilon \nabla^{2} \phi_e = - 4 \pi \rho_e = - 4 \pi \sum_{i=1}^{N} z_i e c_i (x,y,z,t)
\label{poisson}
\end{equation} 
where $\varepsilon$ is the permittivity of the solvent and $\rho_e$ is the density of free charges in the solvent.  Since the Reynolds number is small for flow in micro-capillaries, the fluid flow is described 
by Stokes equation
\begin{equation} 
- \bnabla p + \eta \nabla^{2} {\bf u} - \rho_e \bnabla \phi_e = 0,
\label{stokes}
\end{equation} 
together with the incompressibility constraint 
\begin{equation} 
\bnabla \cdot {\bf u} = 0,
\end{equation} 
where $p$ is  the pressure and 
$\eta$ is the dynamic viscosity of the solvent.

These equations are associated with a number of boundary conditions that 
describe our situation. The concentrations $c_i$ reduce to their equilibrium distributions in a uniform capillary 
very far away from the sample zone 
\begin{equation} 
c_{i} (x,y,z)  \sim  c_{i}^{eq} (y,z) \quad \mbox{if  $|x| \rightarrow \infty$}.
\end{equation} 
Clearly $c_{i}^{eq} (y,z) = 0$ when $i$ corresponds to the sample and is given by solutions of the two dimensional Poisson-Boltzmann model 
for the other species (background electrolytes). The potential $\phi_e$ obeys Dirichlet  boundary conditions 
\begin{equation} 
\phi_e (x,y,z) = \left\{     
\begin{array}{ll}
\phi_{e}^{w} &  \mbox{if $(x,y,z) \: \varepsilon \: S_w$} \\
V &  \mbox{if $(x,y,z) \: \varepsilon \:  S_{+}$} \\
0 &  \mbox{if $(x,y,z) \: \varepsilon \: S_{-}$} 
\label{dirichlet}
\end{array}
\right.
\end{equation}
where $\phi_{e}^{w}$ is the value of the potential at the wall, $V$ is the applied voltage, and, $S_w$, $S_{+}$ and $S_{-}$ are the parts of the 
bounding surface of the capillary that correspond to the walls, the inlet and the outlet respectively (Figure~1). The inlet and the outlet 
sections are separated by a distance $L$ and we are interested in the limit where $V$ and $L$ approach infinity but $V/L = E_0$ 
remains finite. The hydrodynamic flow satisfies the no-slip conditions at the wall
\begin{equation} 
{\bf u} (x,y,z,t) = 0 \quad   \mbox{if $(x,y,z) \: \varepsilon \: S_w$},
\label{no-slip}
\end{equation} 
and,
\begin{equation} 
{\bf u} (x,y,z,t)  \sim  {\bf u}^{eq} (y,z) \quad \mbox{if $|x| \rightarrow \infty$},
\end{equation} 
where ${\bf u}^{eq} (y,z)$ is the electro-osmotic flow in the uniform capillary when no sample is present, and, when $L$ is infinite. 
Finally, the ion fluxes are zero at the capillary walls ($S_w$)
\begin{equation} 
(  z_i e \nu_i c_i \bnabla \phi_e  + D_i \bnabla c_i ) \cdot \hat{ {\bf n} } = 0 \quad \mbox{if $(x,y,z) \: \varepsilon \: S_w$} 
\label{zero_flux}
\end{equation}
where $\hat{ {\bf n} }$ is the unit normal at the wall directed into the fluid.

These equations and boundary conditions form a well posed system of equations that may be integrated numerically 
for given initial conditions. Such a description would, however, be unnecessarily complex. Our objective is to introduce a 
series of approximations that enable a reduced description in terms of a single one dimensional evolution equation 
for the cross-sectionally averaged concentrations $\bar{c}_{i} (x,t)$, since this is the quantity that is essentially measured by 
the detector. Henceforth, a bar above a field variable will always denote its average over the cross-section of the capillary.

\section{One dimensional homogenized equations}
\label{sec:homogenized}
We now introduce a number of approximations that simplify the description and lead up to the one dimensional 
homogenized equation for the sample ion concentration that we seek. The developments closely follow GC 
except for the important difference that the possibility of a hydrodynamic flow field ${\bf u}$ 
is admitted. 

\subsection{Thin Debye layers} 
\label{ssec:TDL}
The fixed charges on the wall are shielded by a cloud of counter-ions over a length scale of the order of the Debye 
length ($\lambda_D$) which is related to the equilibrium  ionic concentrations~\citep{probstein}.
The Debye length is typically of the order of nanometers; much smaller than the characteristic width of the 
micro-channel which is in the range of tens of microns. Under such conditions, the left hand side of (\ref{poisson})
may be set to zero~\citep{Planck}, and,  as a consequence, the last term in (\ref{stokes}),  representing the density of electric forces also 
vanish. Electrical forces could also arise outside of Debye layers in response to variations in electrical 
conductivity~\citep{chen_convective_2005,oddy_multiple-species_2005}. The contribution 
from such forces may however be neglected in the current context (see Supplementary Material online).
Thus, Poisson's equation is replaced by the constraint of local electro-neutrality.  The loss of the equation for $\phi_e$ is however 
not catastrophic, because $\phi_e$ can be determined from the equation of current conservation 
\begin{equation} 
\bnabla \cdot {\bf J}_{e} = 0 
\label{current_conserve}
\end{equation} 
where the current density 
\begin{equation} 
{\bf J}_{e} = - \left( \sum_{i=1}^{N} z_i^{2} e^{2} \nu_{i} c_{i} \right)  \bnabla \phi_e -  \sum_{i=1}^{N} z_{i} e D_{i} \bnabla c_{i} 
\label{Je}
\end{equation} 
is the sum of the Ohmic conduction (first term) and a contribution due to differential diffusion (second term). Equation (\ref{current_conserve}) 
follows from the constraint of local electro-neutrality on summing (\ref{transport}) after multiplying by $z_i e$. The loss of the electrical forcing 
term in (\ref{stokes}) is mitigated by replacing the no-slip boundary condition (\ref{no-slip}) with  the Helmholtz-Smoluchowski slip boundary 
condition~\citep{probstein}
\begin{equation} 
{\bf u} (x,y,z) =   \frac{\varepsilon \zeta \bnabla \phi_e }{4 \pi \eta}  \quad   \mbox{if $(x,y,z) \: \varepsilon \: S_w$} 
\label{HS-slip}
\end{equation} 
that takes into account the presence of a boundary layer of non-vanishing charge next to the capillary wall. Here $\zeta$, the so called 
``zeta-potential'' is the difference in the values of $\phi_{e}$ between a point at the outer edge of the Debye layer and the corresponding point on the wall.
It is a material property that will be assumed constant.  The boundary conditions 
(\ref{dirichlet}) are replaced by
\begin{equation} 
{\bf J}_{e} \cdot \hat{{\bf n}} = 0 
\label{bc_wall_insulator}
\end{equation} 
which follows from (\ref{zero_flux}) and is the condition of zero current flow into capillary walls. At large distances from the sample zone 
we have a uniform field ($E_0$) and a uniform flow ($u_{eo}$), thus,
\begin{equation} 
\phi_e (x,y,z,t) \sim - E_0 x  \quad \mbox{if   $|x| \rightarrow \infty$} 
\end{equation}
and 
\begin{equation} 
{\bf u} (x,y,z,t)  \sim  {\bf u}^{eq} (y,z)  =  - \frac{\varepsilon \zeta E_0 }{4 \pi \eta}  \;  \hat{{\bf x}}= u_{eo}  \: \hat{{\bf x}} \quad \mbox{if   $|x| \rightarrow \infty$} 
\end{equation} 
where $\hat{{\bf x}}$ is the unit vector in the axial direction.

\subsection{Three ion model}
\label{ssec:3ion}
We introduce the Kohlrausch \citep{kohlrausch} function
\begin{equation} 
K(x,y,z,t) = \sum_{i=1}^{N} \frac{1}{\nu_{i}} c_i(x,y,z,t).
\end{equation} 
Then (\ref{transport}) together with the constraint of local electro-neutrality yields the following equation for $K$:
\begin{equation}
\frac{ \partial K }{\partial t} + \bnabla \cdot ( {\bf u}  K ) = \nabla^{2} \left(  \sum_{i=1}^{N} D_i  \frac{c_i}{\nu_i} \right).
\label{transport_K}
\end{equation} 
For a two ion system, equation~(\ref{transport_K}) together with the condition of local electro-neutrality leads to the 
Ohmic model~\citep{Melch_Taylor_ARFM,chen_convective_2005}
where the local electrical conductivity evolves as a passive scalar with an effective diffusivity $D_{e} = 2 D_{+} D_{-} /( D_{+} + D_{-} )$; $D_{+}$ 
and $D_{-}$ denotes the diffusivities of the cation and anion. In our problem, we have more than two ionic species and the Ohmic model is not applicable.
However, if we assume that the ions all have the same mobilities, $\nu_i = \nu$,  and therefore, on account of the Einstein relation $D_i/\nu_i = k_B T$, the 
same diffusivities, $D_i = D$,  then the function $K$, rather than the conductivity, is found to evolve as a passive scalar:
\begin{equation}
\frac{ \partial K }{\partial t} + \bnabla \cdot ( {\bf u}  K ) =  D \nabla^{2} K.
\end{equation} 
We restrict ourselves to a minimal model consisting of a system of just three ions ($N=3$).
We will drop the index $i$ and instead  use the suffix $p$ and $n$ to denote the positive ion and negative ion respectively,
in the background electrolyte.
The absence of a suffix will indicate the sample ion. For example,  $c_p$, $c_n$ and $c$ are the concentrations of positive ions,
negative ions and sample ions respectively. Then from the local electro-neutrality constraint and the definition of $K$
we have 
\begin{equation} 
c_n - c_{n}^{(\infty)} =   \frac{z - z_p}{z_p - z_n}  c +  \frac{ \nu  z_p}{z_p-z_n} \, \delta K,
\label{cn}
\end{equation} 
\begin{equation} 
c_p +  \frac{z_n}{z_p} c_{n}^{(\infty)} = -  \frac{z - z_n}{z_p - z_n}  c -  \frac{ \nu z_n}{z_p-z_n} \, \delta K,
\label{cp}
\end{equation} 
where $c_{n}^{(\infty)}$ is the concentration of negative ions in the background electrolyte far from the sample zone.
If $K_{\infty}$ denotes the value of $K$ in the far field, then the perturbation $\delta K = K - K_{\infty}$ is advected by the flow and spreads diffusively from the injection region.
Therefore (GC), after an initial transient that is small compared to the total analysis time, the sample peak,
which moves by electrophoresis in addition to being advected by the flow, migrates 
into a region of space where $\delta K$ is effectively zero. Thus, in the vicinity of the sample peak, we may assume that 
$\delta K = 0$, so that, the background ion concentrations $c_p$ and $c_n$ 
may be expressed as linear functions of the sample ion concentration $c$. 

The diffusion current represented by the second term in (\ref{Je}) vanishes when $D_i = D$ on account of 
local electro-neutrality, so that it reduces to Ohm's law for a homogeneous electrolyte: ${\bf J}_e = - \sigma_e \bnabla \phi_e$, where 
\begin{equation} 
\sigma_e = \sum_{i=1}^{N} z_i^{2} e^{2} \nu_{i} c_{i} = \sigma_{\infty} ( 1 - \alpha \phi )
\end{equation} 
is the local electrolyte conductivity. The expression on the right is obtained on eliminating $c_p$ and $c_n$ using 
(\ref{cn}) and (\ref{cp}) with $\delta K = 0$. Here $\sigma_{\infty}$ is the bulk electrolyte conductivity,
$\phi = c /c_{n}^{(\infty)}$ is the sample concentration relative to the background 
and $\alpha$ is the parameter 
\begin{equation} 
\alpha = \frac{(z-z_n)(z-z_p)}{z_n (z_p - z_n)}
\end{equation} 
introduced in GC to characterize the nature of the nonlinearity.

\subsection{The lubrication limit}
\label{ssec:lube}
The subsequent development is  based on the premise of  ``slow axial variations''
on account of the inlet to detector separation,  $L$, being much larger than the characteristic channel width, $w_{0}$. 
In practice, $L$ is of the order of  tens of cm whereas $w_{0} \sim 10-100$ $\mu m$. Thus, $L/w_{0} \sim 10^{3} - 10^{4}$.
Therefore, as the sample moves down the micro-channel, the sample concentration 
and all other quantities  controlled by this concentration vary on an axial length scale $L_x \gg w_{0}$.
To see this, suppose that the initial sample concentration was a delta function in the axial direction. 
Then it would spread over a distance  of the order of the channel width in time $\tau_d \sim w_{0}^{2} / D$. 
The total analysis time is  $\tau_a \sim L / v_{0}$, where $v_{0}$ is a characteristic migration 
velocity. Thus, $\tau_a / \tau_d \sim (L/w_{0}) \mbox{Pe}^{-1}$ where $\mbox{Pe} = v_{0} w_{0} /D$ is the P\'{e}clet number.
Since typically $\mbox{Pe} \sim 10 - 100$, $\tau_a / \tau_d \gg 1$ so that concentrations are 
homogenized across the channel and the fluid flow and transport problems both become quasi one dimensional
a short time ($\sim \tau_d$) after sample injection.

The equation of current conservation, (\ref{current_conserve}), may then be integrated using the 
boundary condition (\ref{bc_wall_insulator}) to give 
\begin{equation}
{\bf E}  =   \hat{{\bf x}} \: E_{0}/(1 - \alpha \bar{\phi}) + \cdots,
 \end{equation} 
 where ${\bf E} = - \bnabla \phi_e$ is the electric field and the neglected terms are asymptotically small in the lubrication limit.
 Thus, the field is predominantly in the axial direction 
 and depends on the local sample concentration. The axial variability of the electric field affects the evolution of the sample  concentration in two ways. 
 First, it results in a variable electrophoretic migration velocity $v_{0}/(1 - \alpha \bar{\phi})$ 
 for the sample ions, where $v_{0} = z e \nu  E_{0}$ is the migration velocity of an isolated sample ion. Second, it results in a 
 variable slip velocity for the electro-osmotic flow through the boundary condition (\ref{HS-slip}). 
 
 The hydrodynamic flow 
 field due to such a variable slip velocity may be written down using lubrication theory~\citep{gh_02c}. The velocity is predominantly 
 in the axial direction, ${\bf u} = u \hat{{\bf x}} + \cdots$ and the axial component $u$ may be written as the sum of a 
 mean $\bar{u}$ and a fluctuation about the mean, $\Delta u$, due to the induced pressure gradient. The mean flow $\bar{u}$ 
 is a constant given by 
 \begin{equation}
 \bar{u} = \lim_{L \rightarrow \infty} \frac{1}{L} \int_{0}^{L} u_s (x,t)  \; dx = u_{eo}
 \end{equation} 
where $u_s (x,t)$ is the slip
velocity at the wall. In the presence of concentration gradients, the slip velocity could also have a diffusiophoretic component~\citep{prieve_motion_1984,rica_electrodiffusiophoresis:_2010}. This is however expected to be small 
in the present context  (see Supplementary Material online).
\begin{equation} 
u_{s} (x,t) = u_{eo} / (1 - \alpha \bar{\phi} ) + \cdots,
\end{equation}
 and,
\begin{equation} 
\label{Deltau}
\Delta u \equiv u - \bar{u} =  \left( u_s - u_{eo} \right) \left( 1 - \frac{u_p}{\bar{u}_{p}} \right) = \frac{\alpha \bar{\phi}  \, u_{eo} }{1 - \alpha \bar{\phi}} \left( 1 - \frac{u_p}{\bar{u}_{p}} \right) + \cdots
\end{equation}
to leading order in the lubrication approximation.
Here $u_p$ is a function that depends solely on the cross-sectional shape and represents the flow profile due to a unit pressure gradient.
It is defined by the solution of the equation 
\begin{equation} 
\frac{\partial^{2} u_p}{\partial y^{2}} + \frac{\partial^{2} u_p}{\partial z^{2}} = - 1 
\end{equation} 
with the boundary condition 
\begin{equation} 
u_p (y,z) = 0 \quad   \mbox{if $(x,y,z) \: \varepsilon \: S_w$}.
\end{equation} 
The function $u_p$ admits analytical representation for several cross-sectional shapes.

\subsection{Taylor-Aris limit and macrotransport}
\label{ssec:taylor}
The time evolution of the concentration field of an advected scalar in the limit $\tau_a/\tau_d \gg 1$ was first described by Taylor and Aris 
\citep{taylor,aris} and later applied to a wide variety of problems involving shear induced dispersion~\citep{macrotransport_processes} including dispersion problems in CE~\citep{ghosal_annrev06}.
In this limit, lateral inhomogeneities in the scalar concentration are small $c(x,y,z,t) = \bar{c} (x,t) + \cdots$, so that 
the cross-sectionally averaged concentration 
$\bar{c}$ is advected by the mean flow ${\bar u}$ and undergoes axial diffusion with an effective diffusivity~\citep{pof}
\begin{equation} 
\label{De}
D_e = D -  \frac{\overline{G  u}}{D} 
\end{equation} 
where $G$ satisfies 
\begin{equation} 
\frac{\partial^{2} G}{\partial y^{2}} + \frac{\partial^{2} G}{\partial z^{2}} = \Delta u = u - \bar{u}
\end{equation} 
and the conditions 
\begin{equation} 
\bnabla G  \cdot \hat{{\bf n}} = 0 \quad   \mbox{if $(x,y,z) \: \varepsilon \: S_w$}.
\end{equation}
To remove the indeterminacy of $G$ up to a constant, we further impose the condition $\bar{G}=0$.
 Furthermore, the sample ions are advected with a concentration dependent velocity  $v_{0}/(1 - \alpha \bar{\phi})$. Thus, the 
 equation satisfied by $\bar{\phi}$  is 
 \begin{equation} 
 \frac{\partial \bar{\phi} }{\partial t} + \frac{\partial}{\partial x} \left[  \left( u_{eo} + \frac{v_{0} }{1 - \alpha \bar{\phi} } \right) \bar{\phi} \right] =  \frac{\partial}{\partial x} \left[ 
 \left\{  D + \frac{k u_{eo}^{2} w_{0}^{2}}{D} \left( \frac{\alpha \bar{\phi} }{1 - \alpha \bar{\phi} }\right)^{2}  \right\}  \frac{\partial \bar{\phi} }{\partial x}  \right] 
 \label{generalized_GC}
 \end{equation} 
 where $k$ is a numerical constant depending  solely on the  cross-sectional shape and $w_{0}$ is a characteristic channel width. 
 The expression for the effective diffusivity $D_{e}$ appearing on the right hand side of (\ref{generalized_GC}) follows on 
 evaluating (\ref{De}) using (\ref{Deltau}).
 In particular, for a planar channel of half-width $w_{0}$, a simple calculation shows that $u_p = (w_{0}^{2} - y^{2} )/2$, which leads to  $k=2/105$. 
 Equation (\ref{generalized_GC}) is the generalization of the evolution equation derived in GC to the situation where 
 the channel has a non-zero electro-osmotic slip velocity. An alternative derivation of (\ref{generalized_GC}) using the 
 method of multiple scales is sketched in the Appendix. 
 
  If $ | \alpha | \bar{\phi} \ll 1$, the nonlinear terms in (\ref{generalized_GC}) may be expanded in 
 Taylor series: $(1 - \alpha \bar{\phi})^{-1} = 1 + \alpha \bar{\phi}  + \alpha^{2} \bar{\phi}^{2} + \cdots$, so that, in place  of  Burgers' equation arrived at in GC 
 we get the following equation
 \begin{equation} 
 \frac{\partial \bar{\phi} }{\partial t} + (v_{0} + u_{eo} + 2 \alpha v_{0}  \bar{\phi} + 3 \alpha^{2} v_{0} \bar{\phi}^{2}  ) \;  \frac{\partial \bar{\phi} }{\partial x} = 
  \frac{\partial}{\partial x} \left\{ \left( D +  \frac{k \alpha^{2}  u_{eo}^{2} w_{0}^{2}}{D} {\bar{\phi}}^{2} \right)  \frac{\partial \bar{\phi}}{\partial x}
\right\}
\label{generalized_weak_GC}
\end{equation} 
which shows an amplitude dependent contribution to the diffusivity in addition to a correction to the term $2 \alpha v_{0}  \bar{\phi}  (\partial \bar{\phi}/\partial x)$ representing 
nonlinear wave steepening. Equation (\ref{generalized_GC}), or, its weakly nonlinear version (\ref{generalized_weak_GC}), is our principal result.

\begin{figure}
  \centerline{\includegraphics[width=6in]{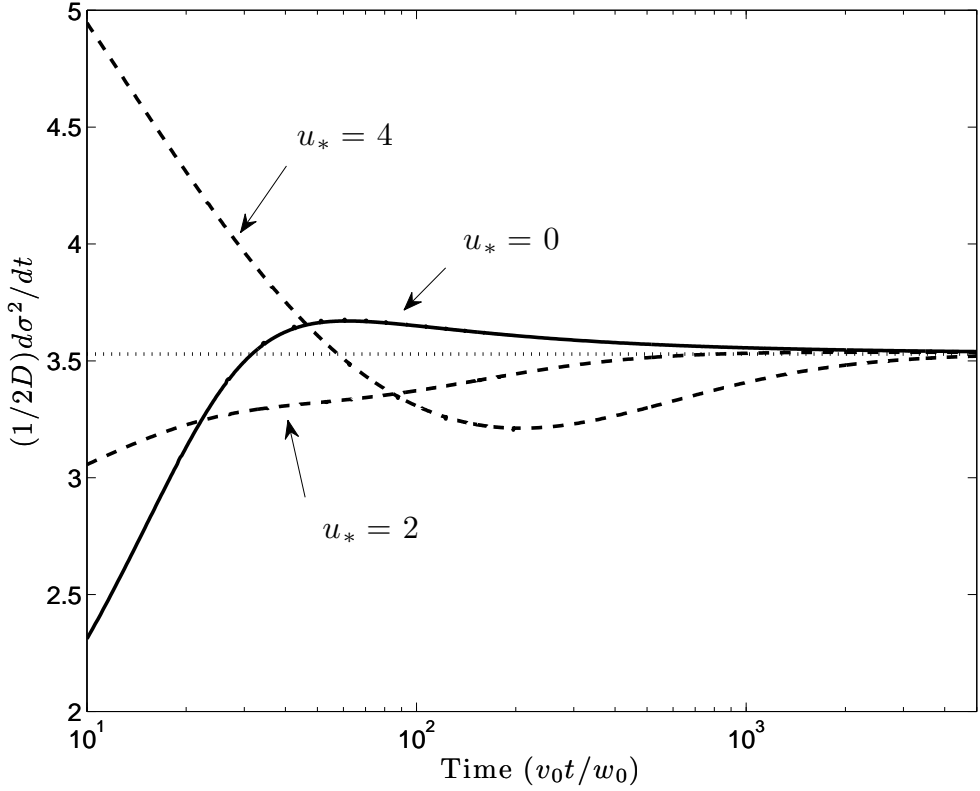}}
  \caption{Time evolution of the normalized rate of increase of variance for three different values of the dimensionless electro-osmotic 
  flow strength $u_{*}=u_{eo}/v_{0}$. Horizontal dotted line is the effective diffusivity predicted 
  by equation~(35) in GC. }
\label{fig:diffusivity}
\end{figure}
Equation (\ref{generalized_GC}) has a singularity at $\bar{\phi} = 1/\alpha$ when 
$\alpha$ is positive. However, (\ref{generalized_GC}) ceases to be valid even before this singularity is reached (GC), 
because, the requirement that $c_p$ and $c_n$ must be non-negative, imposes the constraint $\phi < \phi_{c}$ 
where $\phi_{c} = (z_p - z_n)/(z_p - z)$ when $z < 0$ and $\phi_{c} = - [ z_n (z_p - z_n) / [ z_p (z - z_n) ]$
when $z > 0$. If $\alpha > 0$, then it may be shown that $\phi_{c} < 1 / \alpha$. The physical reason for the 
breakdown of  (\ref{generalized_GC}) when $\phi > \phi_{c}$ is the following: under those conditions the 
conductivity in the sample zone is so high as to reduce the electric field to very small values. Then the electrophoretic 
motion of the sample ions become very small and there is little relative motion between 
the sample peak and $\delta K$. Thus, the assumption that  $\delta K$ may be set to  zero in the sample zone
can no longer be employed and (\ref{generalized_GC}) is no longer valid. However, after sufficient time has passed and axial 
diffusion has caused the peak value of $\bar{\phi}$ to fall below $\phi_c$, the time evolution once again 
proceeds in accordance with (\ref{generalized_GC}). Numerical simulations of the full electrohydrodynamic equations 
described in \S~\ref{ssec:TDL} confirm this behavior (Chen \& Ghosal, unpublished).
For the purpose of numerical integration, (\ref{generalized_weak_GC}) is a little more convenient  than (\ref{generalized_GC}) 
since it does not exhibit the singularity
at $\bar{\phi} = 1/ \alpha$.
Nevertheless, even though solutions to (\ref{generalized_weak_GC}) may be formally calculated even for 
$\bar{\phi} > \phi_c$, the solution in this regime is devoid of physical 
significance, and indeed, will yield negative concentrations of background electrolytes in parts of the domain if (\ref{cn}) and (\ref{cp}) 
with $\delta K=0$ are employed to calculate the concentrations of the background ions.

\section{Dispersion}
\label{sec:dispersion}
Equation (\ref{generalized_GC}) or (\ref{generalized_weak_GC}) provide a compact description of electromigration dispersion that 
is helpful for gaining a qualitative understanding of  the underlying mechanisms. Furthermore, it
is much more amenable to numerical integration  than the full three dimensional coupled problem involving fluid flow 
and transport. By way of example, in this section, we use numerical integration to 
illustrate a point that at first sight may appear counter-intuitive, but, is clarified by an analysis of (\ref{generalized_GC}).

Since the presence of a zeta-potential results in cross-channel variations  in the flow velocity due to 
 induced pressure gradients, one would expect the efficiency of separation to be adversely affected. However, 
 in certain ranges of parameters the reverse may  actually be true. To understand this, one needs to examine the roles of the 
 different terms in (\ref{generalized_weak_GC}). The nonlinear term on the left hand side 
causes wave steepening leading to the formation of shock like structures. This is the 
dominant mechanism that contributes to electromigration dispersion (GC). Taylor dispersion can actually mitigate this tendency by increasing the effective 
axial diffusivity that serves to diffuse the electrokinetic shock. However, on the other hand, if the Taylor dispersion is too large, its contribution to the axial dispersion 
dominates with a consequent loss of separation efficiency. Thus, there is an intermediate value of the wall zeta potential that 
corresponds to the lowest peak dispersion.

\begin{figure}
  \centerline{\includegraphics[width=6in]{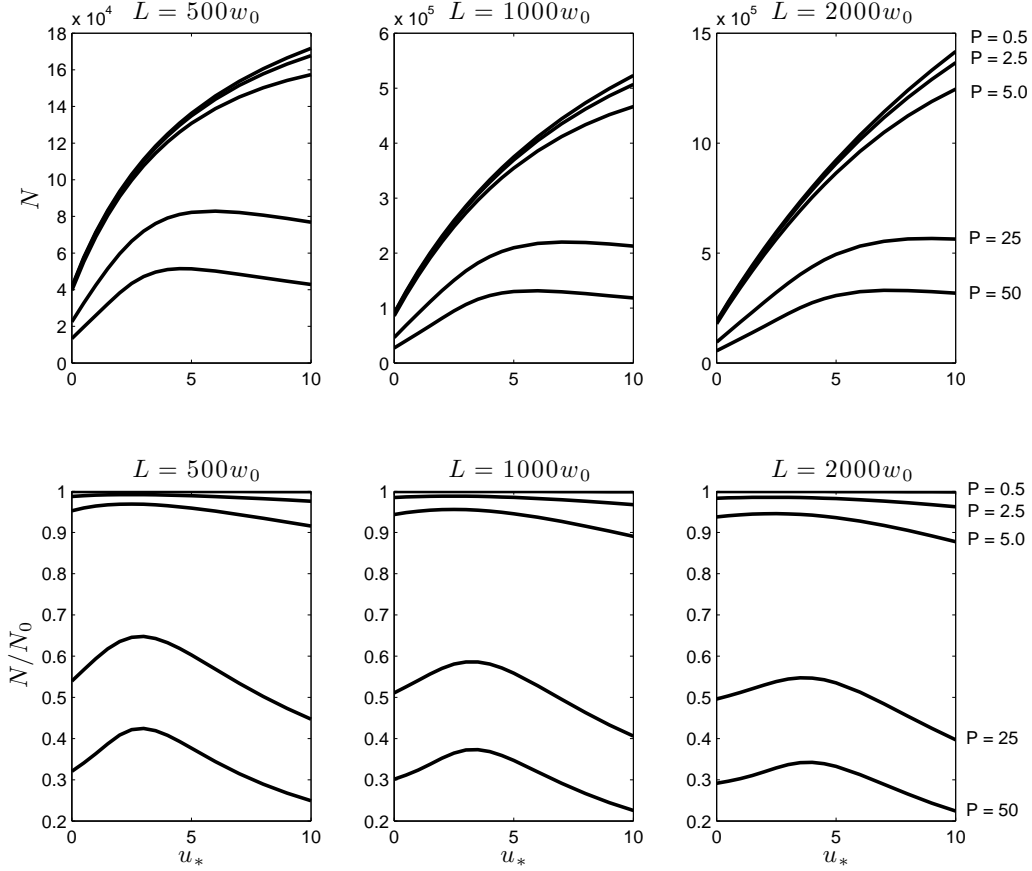}} 
  \caption{The number of theoretical plates $N=L^{2}/\sigma^{2}$ as a function of dimensionless electro-osmotic 
  velocity $u_{*} = u_{eo}/v_{0}$ for three different detector to injection point distances ($L$) and a range of sample loading characterized by the 
  P\'{e}clet number $P = v_{0} \Gamma / D$. Here $N_{0}$ is the ``ideal'' value of $N$ in the 
  absence of electromigration dispersion.}
\label{fig:N}
\end{figure}
Equation (\ref{generalized_GC}) was integrated numerically using a finite volume method that 
allows for adaptive grid refinement and variable time 
steps. We used the ``ode15s'' solver in  MATLAB~\citep{shampine_reichelt97}
(see Supplementary Materials online for further details).
The geometry chosen was that of a planar channel of width $2 w_{0}$.
A ``moving window'' that is advected with the mean migration velocity $v_{0} + u_{eo}$ was  used to 
optimize the number of grid points needed. The size of the window was chosen large enough that  $\bar{\phi}$ can be set to zero at the domain boundaries 
with negligible loss in accuracy. The initial concentration profile $\bar{\phi}(x,0)$ was chosen as a Gaussian of standard deviation $w_{0}$ centered on $x=0$
and with different peak strengths. The ``sample loading''  is 
characterized (GC) by a P\'{e}clet number $P = v_{0} \Gamma / D$ based on the length scale 
\begin{equation} 
\Gamma \equiv \int_{-\infty}^{+\infty} \bar{\phi} \; dx,
\end{equation} 
which is an integral of motion.
A second P\'{e}clet number that characterizes the diffusion, $\Pen = v_{0} w_{0} / D$ is held fixed at the value $200$.
The parameter $\alpha$ was set to $0.5$. These conditions are fairly typical for laboratory experiments.

Figure~\ref{fig:diffusivity} shows the evolution of the quantity $D_{t} \equiv (2D)^{-1} d \sigma^{2} / dt$, where $\sigma^{2}$ is the variance
of $\bar{\phi}(x,t)$, as a function of the 
dimensionless time $v_{0} t / w_{0}$. At late times, when $\bar{\phi}$ is sufficiently small, (\ref{generalized_GC}) may be replaced by (\ref{generalized_weak_GC}).
At even larger times, the quadratic terms in $\bar{\phi}$ become vanishingly small, so that,  the equation 
describing the evolution of $\bar{\phi}$ essentially 
reduces to Burgers' equation as discussed in GC with the minor difference that the constant part of the advection velocity 
 is $v_0 + u_{eo}$ and not $v_0$. 
Therefore, one would expect that $D_t$, which is like an ``instantaneous'' diffusivity,  should asymptote to the value given by equation (35) in GC. 
Figure~\ref{fig:diffusivity} shows that indeed, this is the case for all values of the flow strength, $u_{*} = u_{eo}/v_{0}$. However, the evolution of $D_t$ 
to the common asymptotic value follows different trajectories depending on the strength of the electro-osmotic flow. For larger values of $u_{*}$, $D_{t}$ 
is initially relatively large because of the contribution of the quadratic (and higher order) terms in $\bar{\phi}$ to the effective 
 diffusivity in (\ref{generalized_GC}).
This is because of Taylor dispersion. At intermediate times, $D_t$ is lower than the asymptotic value predicted by equation (35) in GC,
a consequence of the fact that a larger effective axial diffusivity delays the formation of shock like structures that result from nonlinear 
wave steepening. The total variance
is the initial variance ($w_{0}^{2}$) plus the area under the curve $D_t$ from $t=0$ to some time $t=t_{f}$ when the peak arrives at a detector located at $x=L$. This is shown in Figure~2 and 3 in the Supplementary Material.

\begin{figure}
  \centerline{\includegraphics[width=6in]{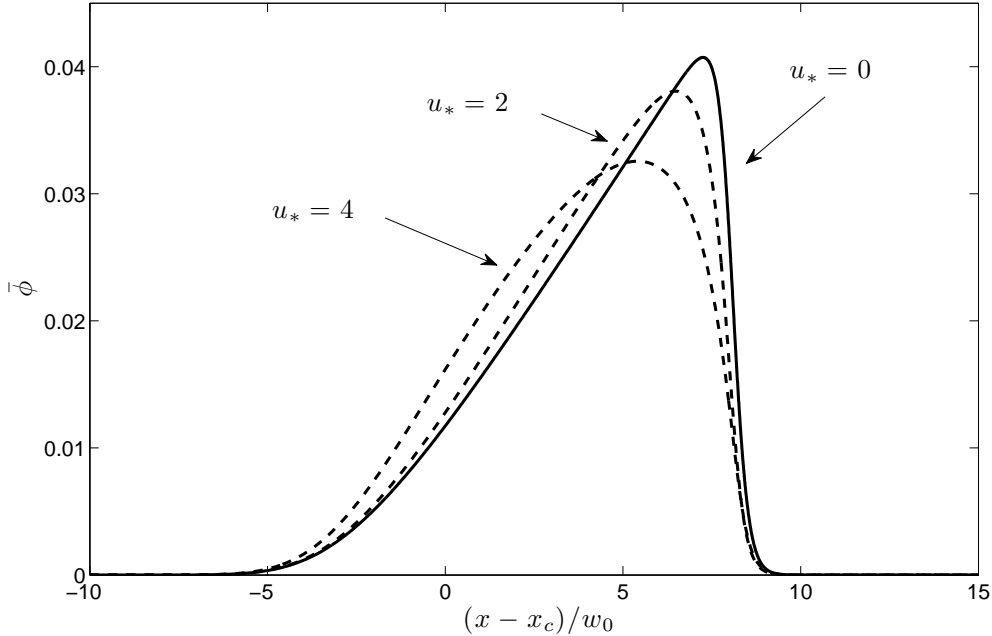}} 
  \caption{Concentration profiles $\bar{\phi}(x,t)$ at a fixed instant of time ($v_{0} t / w_{0} = 200$) for several values of the
  electro-osmotic flow strength, $u_{*}=u_{eo}/v_{0}$. The increased effective axial diffusivity due to Taylor dispersion 
  ``softens'' the electromigration shock that tends to form at the leading edge. Here $P=50$ and  $x=x_{c}$ is the location of the centroid
  of the peak.}
\label{fig:profile}
\end{figure}
The efficiency of separation in CE is often characterized by the ``number of theoretical plates''
$N = L^{2} / \sigma^{2}$ which is a dimensionless measure of the peak width. In the top panel of Figure~\ref{fig:N}, we plot $N$ as a function of $u_{*}$ 
for a number of downstream detector positions and for sample loading ranging from $P=0.5$ (weak) to $P=50$  (strong).  It is clear that increasing $P$ 
reduces $N$ due to electromigration dispersion. However, the dependance on $u_{*}$ is non-monotonic with $N$ increasing at first with $u_{*}$ 
but then decreasing or leveling off to a plateau depending on the location of the detector. Increasing the electro-osmotic flow $u_{eo}$ generally increases 
$N$ due to an obvious and quite trivial reason. If the detector position is fixed, the peak reaches it faster for higher electro-osmotic flow speeds and 
the accumulated variance is less simply because the peak has evolved for a shorter time period. In fact, in the absence of the nonlinearities induced by electromigration 
dispersion, $\sigma^{2} \equiv \sigma_{0}^{2} = w_{0}^{2} + 2 D t_{f} = w_{0}^{2} + 2 D L / (u_{eo} + v_{0} )$. In this 
ideal limit  $N \equiv  N_{0} =  L^{2} / \sigma_{0}^{2}$. In order to eliminate this obvious effect that electro-osmotic flow has 
on peak dispersion, we plot 
$N/N_{0}$ as a function of $u_{*}$ in the lower panel of Figure~\ref{fig:N}, which clearly shows, that the degradation of separation efficiency by electromigration 
dispersion is minimized at an intermediate value of the electro-osmotic flow $u_{*}$.  
The dependence of $N$ on $L$ at different 
values of $P$ is shown in Figure~4 of the Supplementary Material included in the online version.

\begin{figure}
  \centerline{\includegraphics[width=6in]{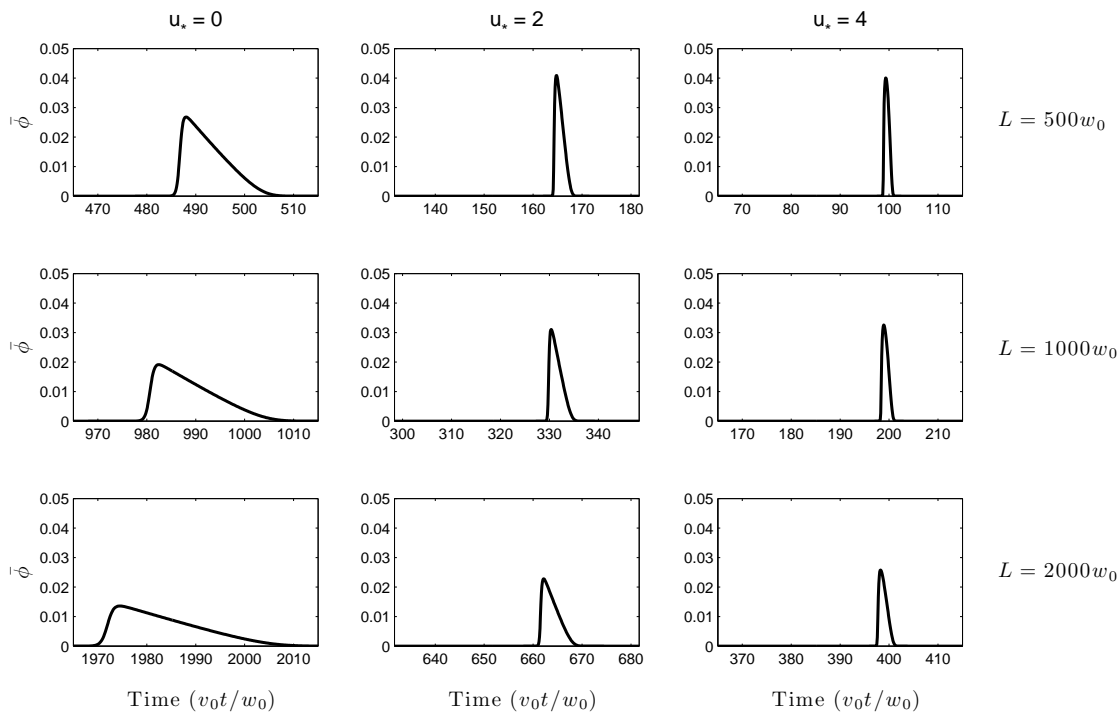} }
  \caption{Concentration $\bar{\phi}(L,t)$ as a function of time of arrival ($t$)  at a fixed detector location ($x=L$)  for several values of 
  electro-osmotic flow strength, $u_{*}=u_{eo}/v_{0}$. It is seen that though electro-osmotic flow improves the resolution, the degree of 
  improvement ``saturates'' as the flow rate is increased, as one might expect from Figure~\ref{fig:N}. Here $P=50$.}
\label{fig:detector}
\end{figure}
The mechanism of this reduction in total variance when $u_{*}$ is 
sufficiently small is evident from Figure~\ref{fig:profile} depicting peak shapes at a fixed time $v_{0} t / w_{0} = 200$ for a number of values of $u_{*}$. 
It is seen that the increase in the effective axial diffusivity due to Taylor dispersion ``softens'' the electrokinetic shock that results from nonlinear 
wave steepening due to electromigration. Thus, electro-osmotic flow has three competing effects (a) reduction in the time available for diffusion
(b) reduction in the nonlinear wave steepening on account of enhanced effective diffusivity  (c) increase in axial diffusion due to the 
Taylor-Aris effect. The total dispersion is determined by the relative contribution of each of (a), (b) and (c). 
 At low values of $u_{*}$, (a) and (b) dominate, but 
at higher values, (c) is more important. 
In laboratory systems, $u_{*}$ could have either sign and a magnitude that could vary between zero (e.g. in a coated capillary where electro-osmosis is suppressed) to 
some number of order unity, since the electro-osmotic  velocity is similar in magnitude to the electrophoretic velocity.
Figure~\ref{fig:detector} provides an alternate viewpoint that has a closer correspondence with 
experiments: the signal intensity $\bar{\phi}(L,t)$ is plotted as a function of the time of arrival at a fixed detector position at distance $L$ 
from the injection point. It is seen that introduction of an electro-osmotic flow at first results in a significant reduction in peak width,
but with diminishing returns as the flow strength is increased.
A similar plot is shown as Figure~6 in the Supplementary Material except there $\bar{\phi}(x,t)$ is plotted as a function 
of $x$ at fixed times. 

\section{Conclusions}
\label{sec:conclusion}
In conclusion, we would like to make a few remarks about the  relevance of the analysis presented here to laboratory practice.
The model studied here is clearly oversimplified and (\ref{generalized_GC}) cannot be expected to predict in detail 
the peak shapes in a real laboratory experiment. In particular, real electrophoresis buffers contain many more than three ions
including one or more weak acids or bases that are added to maintain a stable pH. Nevertheless, it is important to understand 
the qualitative effects predicted by the simplified  model considered here, since the actual dispersion in a real laboratory experiment is likely to be due 
to a multiplicity of  causes, of which, the mechanism discussed here would be one. 

In this paper, we have described the effects of electromigration dispersion by means of a small set of dimensionless 
parameters: the P\'{e}clet numbers $\Pen = v w_{0} / D$, $P = v \Gamma/D$, $\alpha$, and, the dimensionless electro-osmotic 
flow speed $u_{*}=u_{eo}/v_{0}$. A rough idea of typical values of these parameters in laboratory experiments 
is helpful. If we take $v_{0} \sim 1$ mm/s, $w_{0} \sim  10 \mu m$, $D \sim 10^{-5}$ cm$^{2}$/s as fairly typical, we have $\Pen \sim 10$.
For large molecules, $D$ could be  an order of magnitude smaller, so that $\Pen$ could be in the hundreds. 
If the sample concentration is of the same order as the background electrolyte concentration and the peak width 
is of the order of the channel width, then $\Gamma \sim w_{0}$. 
However, such sample concentrations would be considered extremely high in CE and would show very strong 
peak distortion. We may therefore take this as an upper limit for $\Gamma$. Thus, $P$ could range from 
essentially zero (the linear regime) to $P  = v \Gamma/D  \sim  v w_{0} /D = \Pen$. Since electrophoresis and electro-osmosis are generally of comparable 
magnitudes, $u_{eo} \sim v$. Thus, $u_{*} \sim 1$, though $u_{*}$ could be close to zero if the channel walls 
have a low zeta potential (as in a coated capillary). The magnitude of the parameter $\alpha$ is of order unity if the sample valence 
and the valence of the background electrolytes are similar. However, for macro-ions $|z| \sim 10$ or larger. For such large value of $z$,
$\alpha \sim z^{2}$ could be in the hundreds, and, nonlinear effects of the kind considered in this paper would be very pronounced. 

Electro-osmotic flow is generally desirable in the context of CE 
as it shortens the transit time from inlet to detector and as a consequence also reduces the peak variance.
The effect of electro-osmotic flow on separation efficiency is however complicated in the presence of electromigration 
dispersion since  a number of competing effects are simultaneously present. This paper generalizes the analysis presented 
in an earlier paper (GC) where the effect of electromigration dispersion in the absence of 
any bulk flow was considered. However, capillary electrophoresis channels are usually made of fused 
silica which has a large negative zeta potential and sustains strong electro-osmotic flow. Thus, the theory 
presented here broadens the scope of the earlier paper making it more relevant to practical systems.

In summary, a homogenized equation for the cross-sectionally averaged concentration of sample ions was derived by exploiting the approximations 
outlined in \S~\ref{sec:homogenized}, which are:
(a) the Debye length is much smaller than the typical channel width 
(b) the electrolyte contains only three ionic species  of equal diffusivity which are also strong electrolytes
(c) the time between injection and detection is much longer than characteristic transport times
 across the capillary. The outcome of the analysis is the replacement of 
the molecular diffusion coefficient in the one dimensional transport equation derived in GC by an effective axial diffusivity  which is a nonlinear 
function of the concentration. If the concentration is not too high, this effective diffusivity is a quadratic function of the 
concentration.

The reduced one-dimensional equation (\ref{generalized_GC})
provides a compact description of electromigration dispersion that 
is much more amenable to numerical integration  than the full coupled three dimensional problem involving fluid flow 
and transport in the capillary. In fact, since both the length to width aspect ratio of the capillary as well as the P\'{e}clet number ($\Pen$)
are usually large in experiments, such  three dimensional simulations can be computationally intensive. The loss of accuracy in replacing the full equations by the one dimensional model is minimal because the requirements for its validity are well satisfied and the 
asymptotic convergence of the thin Debye layer and long length scale approximations have been shown to be rapid
\citep{ghosal_annrev06,loc,pof}. 

The effect of a wall zeta potential may be explained qualitatively in the following way: alteration of the ionic composition of the solute 
in the sample zone results in a change of the electrical conductivity and therefore in the axial electric field. Thus, the electro-osmotic slip velocity 
now has an axial variation which induces pressure gradients along the channel. This in turn induces shear in the electro-osmotic 
velocity profile.  The consequent shear induced (Taylor-Aris) dispersion results in a concentration dependent axial 
diffusivity. Numerical integration of the homogenized equation (\ref{generalized_GC}) provided some qualitative insights into the 
effect of this nonlinear diffusion on separation efficiency. It is seen that the total variance 
for a fixed detector position actually decreases with the strength of the electro-osmotic flow as long as the flow is not too strong. 
This is due to the fact that an increase in the effective diffusivity for axial transport counteracts the wave steepening resulting from 
electromigration dispersion, and, furthermore, the transit time from injection to detection is reduced.
 However, if the flow is much stronger, the dispersion caused by the enhanced  axial diffusivity itself 
is dominant, and this degrades any gain in resolution due to the aforemetioned causes.
 Thus, if all other parameters are invariant, the peak variance is a non-monotonic 
function of the electro-osmotic flow strength and is  optimal at a certain intermediate value of the flow velocity.

\vspace*{0.1in}
This work was supported by the NIH under grant R01EB007596


\appendix
\section{Derivation of (\ref{generalized_GC}) by the method of multiple scales}\label{appendix}
\noindent It is advantageous to work with dimensionless variables. This is achieved by using $w_{0}$ as the unit of length and $w_{0}/v_{0}$ as the 
unit of time. The governing equations are therefore those presented in \S~\ref{sec:formulation} and \S~\ref{sec:homogenized}
but with $w_{0}$ and $v_{0}$ set to one. These equations are
\begin{eqnarray}
[(1 - \alpha \phi) X]_{x} + [(1 - \alpha \phi) Y]_{y} +  [(1 - \alpha \phi) Z]_{z} &=& 0  \label{divj}\\
\phi_{t} + [(u+X)\phi]_{x} +  [(v+Y)\phi]_{y} +  [(w+Z)\phi]_{z} &=& \Pen^{-1} [ \phi_{xx} + \phi_{yy} + \phi_{zz} ] \label{transportA}\\
- p_{x} + u_{xx} + u_{yy} + u_{zz} &=& 0 \label{stokesx}\\
- p_{y} + v_{xx} + v_{yy}  + v_{zz}  &=& 0 \label{stokesy}\\
- p_{z} + w_{xx} + w_{yy} + w_{zz} &=& 0 \label{stokesz}
\end{eqnarray}
where $(X,Y,Z)$ is the electric field vector (normalized by $E_{0}$), ${\bf u} = (u,v,w)$, $\phi$ is the normalized sample concentration, $\Pen = v_{0} w_{0}/D$ (P\'{e}clet number) and suffixes denote partial derivatives. 
The flow is solenoidal:
\begin{equation} 
\label{divu0}
u_x + v_y + w_z = 0
\end{equation} 
and the electric field is irrotational:
\begin{equation} 
\label{irrotational}
X_y = Y_x \quad X_z = Z_x  \quad Y_z = Z_y 
\end{equation} 
The boundary conditions at the wall are 
\begin{equation}
\label{HSslip}
u = u_{*} X \quad v = 0 \quad w = 0
\end{equation} 
and 
\begin{eqnarray} 
m \phi_y + n \phi_z &=& 0 \label{zeroflux}\\
m Y + n  Z &=& 0 \label{zerocurrent}
\end{eqnarray} 
where $(0,m,n)$ is the unit normal on the wall and $u_{*} = u_{eo}/v_{0}$. At points far upstream and downstream of the sample peak 
\begin{equation} 
\label{farbc}
X \sim 1 \quad u \sim u_{*} \quad v \sim 0 \quad w \sim 0 \quad \phi \sim 0.
\end{equation} 
This description presumes infinitely thin Debye layers (\S~\ref{ssec:TDL}) and constancy of the Kohlrausch 
function in a three ion system  (\S~\ref{ssec:3ion}).

We now assume that all axial variations happen on a slow spatial scale for which we introduce the slow 
variable 
$\xi = \epsilon x$
and the corresponding slow times $\tau = \epsilon t$ and $T = \epsilon^{2}t$, 
where $\epsilon$ is a small parameter. The physical origin of the two times is due to the existence of the 
time-scale $w_0 / v_0$ which is the advective time over a channel diameter and the diffusion 
time-scale  $w_{0}^{2} / D$ for diffusion  across the channel diameter. Since the flow and fields are 
predominantly axial, we introduce 
\begin{equation} 
\tilde{Y} = Y/\epsilon \quad \tilde{Z} = Z/\epsilon \quad \tilde{v} = v /\epsilon \quad \tilde{w} = w /\epsilon \quad \tilde{p} = \epsilon \, p
\end{equation} 
where the pressure scaling is designed to retain the pressure gradient term at leading order. 
Next we re-write (\ref{divj})-(\ref{stokesz}) in terms of the scaled variables using the transformations $\partial_t = \epsilon \partial_{\tau} + \epsilon^{2} \partial_{T}$ 
and $\partial_{x} = \epsilon \partial_{\xi}$ and expand all dependent variables $f$ in asymptotic series $f = f^{(0)} + \epsilon f^{(1)} + \epsilon^{2} f^{(2)} + \cdots$ 
Equating like powers of $\epsilon$ on either side of the governing equations we get a series of successive problems to solve.

\subsection{Zeroth Order} 
From  (\ref{divj}) and (\ref{irrotational}) 
\begin{equation} 
\label{divj0}
[(1 - \alpha \phi^{(0)} ) X^{(0)} ]_{\xi} + [(1 - \alpha \phi^{(0)}) \tilde{Y}^{(0)} ]_{y} +  [(1 - \alpha \phi^{(0)} ) \tilde{Z}^{(0)} ]_{z} = 0,
\end{equation}
\begin{equation} 
X^{(0)}_{y} = X^{(0)}_{z} = 0. 
\end{equation} 
Thus, $X^{(0)}$ is independent of $y$ and $z$. Integrating (\ref{divj0}) over the channel cross-section and using the boundary condition (\ref{zerocurrent}) 
we get 
\begin{equation} 
[ (1 - \alpha \bar{\phi}^{(0)} ) X^{(0)} ]_{\xi} = 0,
\end{equation} 
and thus, on account of the far field conditions (\ref{farbc}) 
\begin{equation} 
\label{X0}
 X^{(0)} = 1 / (1 - \alpha \bar{\phi}^{(0)} ).
 \end{equation} 
 Equation (\ref{transportA}) gives at lowest order 
 \begin{equation} 
  \phi^{(0)}_{yy} + \phi^{(0)}_{zz} = 0,
  \end{equation} 
  which, with the Neuman boundary condition (\ref{zeroflux}) implies that $\phi^{(0)}$ is independent of $y$ and $z$, thus,
\begin{equation} 
\phi^{(0)} = \bar{\phi}^{(0)} (\xi,\tau,T).
\end{equation} 
Finally, the flow equations (\ref{stokesx})-(\ref{stokesz}) give 
\begin{equation}
- \tilde{p}^{(0)}_{\xi} + u^{(0)}_{yy} + u^{(0)}_{zz}  = 0, \label{stokesx0}
\end{equation}
\begin{equation}
- \tilde{p}^{(0)}_{y} = - \tilde{p}^{(0)}_{z} = 0.
\end{equation} 
Thus, $\tilde{p}^{(0)}$ is independent of $y$ and $z$. Now  (\ref{stokesx0}) with boundary condition derived from (\ref{HSslip}) may be readily 
integrated, 
\begin{equation} 
\label{u0}
u^{(0)} = - \tilde{p}^{(0)}_{\xi} u_{p} + u_{*} X^{(0)} = - \tilde{p}^{(0)}_{\xi} u_{p} + u_{*} / (1 - \alpha \bar{\phi}^{(0)} ).
\end{equation} 
where $u_p$ is the solution of  $(u_p)_{yy} + (u_p)_{zz} = -1 $ that vanishes at the wall.  To determine the unknown function
$\tilde{p}^{(0)}$ we use (\ref{divu0}). Integrating over the cross-section and using (\ref{HSslip}) we get 
\begin{equation}
\bar{u}^{(0)} = - \tilde{p}^{(0)}_{\xi} \bar{u}_{p} + u_{*} / (1 - \alpha \bar{\phi}^{(0)} ) = u_{*},
\end{equation}
which determines $\tilde{p}^{(0)}_{\xi}$. Substituting this in (\ref{u0}) we obtain the axial flow velocity 
which is identical to the lubrication theory solution \citep{gh_02c} applied to the case of an axially varying slip 
velocity $u_{*} / (1 - \alpha \bar{\phi}^{(0)} )$ along the channel wall:
\begin{equation}
\label{u00}
u^{(0)} = \frac{u_{*}}{1 - \alpha \bar{\phi}^{(0)}} \left[ 1 - \frac{\alpha \bar{\phi}^{(0)}}{\bar{u}_{p}} u_{p} (y,z) \right].
\end{equation}
Since $\phi^{(0)} = \bar{\phi}^{(0)}$, (\ref{irrotational}) and (\ref{divj0}) show that the two dimensional vector $(\tilde{Y}^{(0)},\tilde{Z}^{(0)})$ 
is both solenoidal and irrotational in the $y-z$ plane, and, on account of the zero flux condition, must within the domain. Therefore,
\begin{equation}
\label{Y0Z0}
\tilde{Y}^{(0)} = \tilde{Z}^{(0)} = 0.
\end{equation} 
\subsection{First Order} 
From (\ref{transportA}) 
\begin{eqnarray}
\label{eq4phi1}
\Pen^{-1} [ \phi^{(1)}_{yy} + \phi^{(1)}_{zz} ]   &=&    \phi^{(0)}_{\tau} + [ (u^{(0)}+X^{(0)}) \phi^{(0)} ]_{\xi} \nonumber \\
& & +  [ (\tilde{v}^{(0)}+\tilde{Y}^{(0)}) \phi^{(0)} ]_{y}  +  [ (\tilde{w}^{(0)} +\tilde{Z}^{(0)}) \phi^{(0)} ]_{z}.
\label{transport1}
\end{eqnarray}
Integrating (\ref{transport1}) over the cross-section of the capillary and using the boundary conditions (\ref{HSslip}) 
and (\ref{zerocurrent}) we get a solvability condition for (\ref{eq4phi1}) 
\begin{equation} 
\bar{\phi}^{(0)}_{\tau} + [ \overline{ (u^{(0)}+X^{(0)}) \phi^{(0)} } ]_{\xi} = 0 
\end{equation} 
which reduces to 
\begin{equation} 
\label{eq4phi_0}
\bar{\phi}^{(0)}_{\tau} + [  \{ u_{*} + (1 - \alpha \bar{\phi}^{(0)} )^{-1} \} \bar{\phi}^{(0)}  ]_{\xi} = 0
\end{equation} 
on using (\ref{X0}) and (\ref{u00}). On using the solvability equation (\ref{eq4phi_0}) in (\ref{eq4phi1}) we get 
\begin{equation}
\label{eq4phi1a}
\Pen^{-1} [ \phi^{(1)}_{yy} + \phi^{(1)}_{zz} ]   = 
 [ (u^{(0)}+X^{(0)}  - \bar{u}^{(0)} - \bar{X}^{(0)} ) \bar{\phi}^{(0)} ]_{\xi}  - u^{(0)}_{\xi} \bar{\phi}^{(0)}
\end{equation}
where we have used the continuity equation (\ref{divu0}) to replace $\tilde{v}^{(0)}_{y} + \tilde{w}^{(0)}_{z}$ with $- u^{(0)}_{\xi}$.
Equation (\ref{eq4phi1a}) is readily integrated. If the constant of integration is expressed in terms of $\bar{\phi}^{(1)}$ 
we get 
\begin{equation} 
\phi^{(1)} = \bar{\phi}^{(1)} - \alpha \, \Pen \, u_{*} G \bar{\phi}^{(0)} \bar{\phi}^{(0)}_{\xi} / [\, \bar{u}_{p} \, ( 1 - \alpha \bar{\phi}^{(0)} ) \,]
\end{equation} 
where $G$ is defined by the equation $G_{yy} + G_{zz} = u_{p} - \bar{u}_{p}$ with the condition that $G$ 
has zero flux at the walls and is normalized so that $\overline{G}=0$.  From (\ref{divj}) 
\begin{equation} 
[(1 - \alpha \phi^{(0)} ) \tilde{Y}^{(1)} ]_{y} +  [(1 - \alpha \phi^{(0)} ) \tilde{Z}^{(1)} ]_{z} = \alpha [ \phi^{(1)} X^{(0)} ]_{\xi}
- [ (1 - \alpha \phi^{(0)} ) X^{(1)} ]_{\xi}
\end{equation} 
where we have used (\ref{Y0Z0}). Integrating over the cross-section and using the boundary condition gives the 
solvability condition 
\begin{equation} 
\alpha [ \bar{\phi}^{(1)} \bar{X}^{(0)} ]_{\xi} - [ (1 - \alpha \bar{\phi}^{(0)} ) \bar{X}^{(1)} ]_{\xi} = 0,
\end{equation}
which may be integrated with the far field boundary conditions to give 
\begin{equation} 
\label{X1}
\bar{X}^{(1)} = \frac{\alpha \bar{\phi}^{(1)}}{(1 - \alpha \bar{\phi}^{(0)} )^{2}}
\end{equation} 

\subsection{Second Order} 
From (\ref{transportA}) 
\begin{eqnarray}
& &  \Pen^{-1} [ \phi^{(2)}_{yy} + \phi^{(2)}_{zz}  ] - [(\tilde{v}^{(0)}+\tilde{Y}^{(0)})\phi^{(1)}]_{y} -  [(\tilde{v}^{(1)}+\tilde{Y}^{(1)})\phi^{(0)}]_{y} \nonumber \\
& &  - [(\tilde{w}^{(0)}+\tilde{Z}^{(0)})\phi^{(1)}]_{z} - [(\tilde{w}^{(1)}+\tilde{Z}^{(1)})\phi^{(0)}]_{z}   = -  \Pen^{-1}  \phi^{(0)}_{\xi \xi} \nonumber \\
& & + \phi^{(1)}_{\tau} + \phi^{(0)}_{T} + [(u^{(0)}+X^{(0)}) \phi^{(1)}]_{\xi} + [(u^{(1)}+X^{(1)}) \phi^{(0)}]_{\xi}  
\end{eqnarray}
Integrating over the cross-section and using the boundary conditions we derive
\begin{equation} 
\label{phi1}
 \bar{\phi}^{(1)}_{\tau} + \bar{\phi}^{(0)}_{T} + [ \overline{u^{(0)} \phi^{(1)}} + \overline{X^{(0)}  \phi^{(1)}} + \overline{X^{(1)} \phi^{(0)}}]_{\xi}  
 = \Pen^{-1} \bar{\phi}^{(0)}_{\xi \xi}.
 \end{equation}
 
\subsection{Reconstitution}
On combining (\ref{phi1}) with (\ref{eq4phi_0}) after substituting for the lower order terms from (\ref{X0}), (\ref{u00}) and (\ref{X1}) 
we get 
\begin{eqnarray} 
& & \epsilon \bar{\phi}^{(0)}_{\tau} + \epsilon^{2} \bar{\phi}^{(1)}_{\tau} + \epsilon^{2} \bar{\phi}^{(0)}_{T} + \epsilon \left[ \left(u_{*} + \frac{1}{(1 - \alpha \bar{\phi}^{(0)})} \right) \bar{\phi}^{(0)} 
 + \epsilon \alpha^{2} k \Pen \:  u_{*}^{2} \frac{(\bar{\phi}^{(0)})^{2} \bar{\phi}^{(0)}_{\xi} }{(1 - \alpha \bar{\phi}^{(0)} )^{2}}  \right. \nonumber \\
 & & \left. + \epsilon u_{*} \bar{\phi}^{(1)}  + \epsilon \frac{\bar{\phi}^{(1)}}{1 - \alpha \bar{\phi}^{(0)}} + \epsilon \alpha \frac{\bar{\phi}^{(0)} \bar{\phi}^{(1)}}{(1 - \alpha \bar{\phi}^{(0)} )^{2}} 
\right]_{\xi} - \epsilon^{2} \Pen^{-1} \bar{\phi}^{(0)}_{\xi \xi}  = 0,
\end{eqnarray}
which may be put in an alternate form that is asymptotically equivalent to it for all terms up to order $\epsilon^{2}$:
\begin{equation} 
\bar{\phi}_{t} + \left[ \left( u_{*} + \frac{1}{1 - \alpha \bar{\phi}} \right) \bar{\phi} - \alpha^{2} k \Pen \: u_{*}^{2} \frac{\bar{\phi}^{2} \bar{\phi}_{x}}{(1 - \alpha \bar{\phi})^{2}}  \right]_{x} - \Pen^{-1} \bar{\phi}_{xx} + O(\epsilon^{3}) = 0.
\end{equation} 
Here $k = - \overline{G u_p} / \bar{u}_{p}^{2}$ is a constant that depends solely on the geometry of the channel cross-section. On transforming back to 
dimensional variables (\ref{generalized_GC}) follows.
\end{document}